   \documentstyle[12pt]{article}
\if@twoside\oddsidemargin25mm
\else\oddsidemargin25mm\evensidemargin25mm\marginparwidth25mm\fi%
\begin{document}
\newcommand{\ft}[2]{{\textstyle\frac{#1}{#2}}}
\newcommand{\QED}{{\hspace*{\fill}\rule{2mm}{2mm}\linebreak}}
\def\dop{{\rm d}\hskip -1pt}
\def\bfone{\relax{\rm 1\kern-.35em 1}}
\def\bfzero{\relax{\rm I\kern-.18em 0}}
\def\inbar{\vrule height1.5ex width.4pt depth0pt}
\def\IC{\relax\,\hbox{$\inbar\kern-.3em{\rm C}$}}
\def\ID{\relax{\rm I\kern-.18em D}}
\def\IF{\relax{\rm I\kern-.18em F}}
\def\IK{\relax{\rm I\kern-.18em K}}
\def\IH{\relax{\rm I\kern-.18em H}}
\def\II{\relax{\rm I\kern-.17em I}}
\def\IN{\relax{\rm I\kern-.18em N}}
\def\IP{\relax{\rm I\kern-.18em P}}
\def\IQ{\relax\,\hbox{$\inbar\kern-.3em{\rm Q}$}}
\def\IR{\relax{\rm I\kern-.18em R}}
\def\IG{\relax\,\hbox{$\inbar\kern-.3em{\rm G}$}}
\font\cmss=cmss10 \font\cmsss=cmss10 at 7pt
\def\ZZ{\relax\ifmmode\mathchoice
{\hbox{\cmss Z\kern-.4em Z}}{\hbox{\cmss Z\kern-.4em Z}}
{\lower.9pt\hbox{\cmsss Z\kern-.4em Z}}
{\lower1.2pt\hbox{\cmsss Z\kern-.4em Z}}\else{\cmss Z\kern-.4em
Z}\fi}
\def\a{\alpha} \def\b{\beta} \def\d{\delta}
\def\e{\epsilon} \def\c{\gamma}
\def\G{\Gamma} \def\l{\lambda}
\def\L{\Lambda} \def\s{\sigma}
\def\cA{{\cal A}} \def\cB{{\cal B}}
\def\cC{{\cal C}} \def\cD{{\cal D}}
\def\cF{{\cal F}} \def\cG{{\cal G}}
\def\cH{{\cal H}} \def\cI{{\cal I}}
\def\cJ{{\cal J}} \def\cK{{\cal K}}
\def\cL{{\cal L}} \def\cM{{\cal M}}
\def\cN{{\cal N}} \def\cO{{\cal O}}
\def\cP{{\cal P}} \def\cQ{{\cal Q}}
\def\cR{{\cal R}} \def\cV{{\cal V}}\def\cW{{\cal W}}
%
%
%
\def\crr{\crcr\noalign{\vskip {8.3333pt}}}
\def\tilde{\widetilde}
\def\bar{\overline}
\def\us#1{\underline{#1}}
\let\shat=\hat
\def\hat{\widehat}
\def\hyp{\vrule height 2.3pt width 2.5pt depth -1.5pt}
\def\square{\mbox{.08}{.08}}
\def\Coeff#1#2{{#1\over #2}}
\def\Coe#1.#2.{{#1\over #2}}
\def\coeff#1#2{\relax{\textstyle {#1 \over #2}}\displaystyle}
\def\coe#1.#2.{\relax{\textstyle {#1 \over #2}}\displaystyle}
\def\half{{1 \over 2}}
\def\shalf{\relax{\textstyle {1 \over 2}}\displaystyle}
\def\dag#1{#1\!\!\!/\,\,\,}
\def\to{\rightarrow}
\def\notin{\hbox{{$\in$}\kern-.51em\hbox{/}}}
\def\shdot{\!\cdot\!}
\def\ket#1{\,\big|\,#1\,\big>\,}
\def\bra#1{\,\big<\,#1\,\big|\,}
\def\equaltop#1{\mathrel{\mathop=^{#1}}}
\def\Trbel#1{\mathop{{\rm Tr}}_{#1}}
\def\inserteq#1{\noalign{\vskip-.2truecm\hbox{#1\hfil}
\vskip-.2cm}}
\def\attac#1{\Bigl\vert
{\phantom{X}\atop{{\rm\scriptstyle #1}}\phantom{X}}}
\def\exx#1{e^{{\displaystyle #1}}}
\def\del{\partial}
\def\delbar{\bar\partial}
\def\nex#1{$N\!=\!#1$}
\def\dex#1{$d\!=\!#1$}
\def\cex#1{$c\!=\!#1$}
\def\eg{{\it e.g.}} \def\ie{{\it i.e.}}
%
\def\cS{{\cal K}}
\def\IE{\relax{{\rm I\kern-.18em E}}}
\def\cE{{\cal E}}
\def\rt{{\cR^{(3)}}}
\def\IGam{\relax{{\rm I}\kern-.18em \Gamma}}
\def\IGa{\IA}
\def\LG{Lan\-dau-Ginz\-burg\ }
\def\cV{{\cal V}}
\def\Rt{{\cal R}^{(3)}}
\def\wabc{W_{abc}}
\def\WABC{W_{\a\b\c}}
\def\W{{\cal W}}
\def\tft#1{\langle\langle\,#1\,\rangle\rangle}
\def\IA{\relax{\hbox{{\rm A}\kern-.82em {\rm A}}}}
\let\picfuc=\fp
\def\hata{{\shat\a}}
\def\hatb{{\shat\b}}
\def\hatA{{\shat A}}
\def\hatB{{\shat B}}
\def\bv{{\bf V}}
\def\spg{special geometry}
\def\sc{SCFT}
\def\leel{low energy effective Lagrangian}
\def\pf{Picard--Fuchs}
\def\pfS{Picard--Fuchs system}
\def\el{effective Lagrangian}
\def\Fb{\overline{F}}
\def\nablab{\overline{\nabla}}
\def\Ub{\overline{U}}
\def\Db{\overline{D}}
\def\zb{\overline{z}}
\def\eb{\overline{e}}
\def\fb{\overline{f}}
\def\tb{\overline{t}}
\def\Xb{\overline{X}}
\def\Vb{\overline{V}}
\def\Cb{\overline{C}}
\def\Sb{\overline{S}}
\def\delb{\overline{\del}}
\def\Gammab{\overline{\Gamma}}
\def\Ab{\overline{A}}
\def\Anh{A^{\rm nh}}
\def\alphab{\bar{\alpha}}
\def\cy{Calabi--Yau}
\def\cabg{C_{\alpha\beta\gamma}}
\def\B{\Sigma}
\def\Bh{\hat \Sigma}
\def\Kh{\hat{K}}
\def\Knh{{\cal K}}
\def\A{\Lambda}
\def\Ah{\hat \Lambda}
\def\R{\hat{R}}
\def\V{{V}}
\def\T{T}
\def\Gammah{\hat{\Gamma}}
\def\twot{$(2,2)$}
\def\K{K\"ahler}
\def\rat{({\theta_2 \over \theta_1})}
\def\lv{{\bf \omega}}
\def\w{w}
\def\CP{C\!P}
\def\o#1#2{{{#1}\over{#2}}}
\newcommand{\be}{\begin{equation}}
\newcommand{\ee}{\end{equation}}
\newcommand{\ba}{\begin{eqnarray}}
\newcommand{\ea}{\end{eqnarray}}
\newtheorem{definizione}{Definition}[section]
\newcommand{\bd}{\begin{definizione}}
\newcommand{\ed}{\end{definizione}}
\newtheorem{teorema}{Theorem}[section]
\newcommand{\bth}{\begin{teorema}}
\newcommand{\eth}{\end{teorema}}
\newtheorem{lemma}{Lemma}[section]
\newcommand{\blem}{\begin{lemma}}
\newcommand{\elem}{\end{lemma}}
\newcommand{\brr}{\begin{array}}
\newcommand{\err}{\end{array}}
\newcommand{\nn}{\nonumber}
\newtheorem{corollario}{Corollary}[section]
\newcommand{\bcorol}{\begin{corollario}}
\newcommand{\ecorol}{\end{corollario}}
\def\twomat#1#2#3#4{\left(\begin{array}{cc}
 {#1}&{#2}\\ {#3}&{#4}\\
\end{array}
\right)}
\def\twovec#1#2{\left(\begin{array}{c}
{#1}\\ {#2}\\
\end{array}
\right)}
\begin{titlepage}
\hskip 5.5cm
\vbox{\hbox{POLFIS-TH 08/96}
\hbox{CERN-TH/96--202}
}
\hskip 1.5cm
\vbox{\hbox{hep-th/9608015}\hbox{August, 1996}}
\vfill
\begin{center}
{\LARGE { Central Extension of Extended Supergravities
in Diverse Dimensions}}\\
\vskip 1.5cm
  {\bf Laura Andrianopoli$^1$,
Riccardo D'Auria$^2$,
Sergio Ferrara$^3$ } \\
\vskip 0.5cm
{\small
$^1$ Dipartimento di Fisica, Universit\'a di Genova, via Dodecaneso 33,
I-16146 Genova\\
and Istituto Nazionale di Fisica Nucleare (INFN) - Sezione di Genova, Italy\\
\vspace{6pt}
$^2$ Dipartimento di Fisica, Politecnico di Torino,\\
 Corso Duca degli Abruzzi 24, I-10129 Torino\\
and Istituto Nazionale di Fisica Nucleare (INFN) - Sezione di Torino, Italy\\
\vspace{6pt}
$^3$ CERN Theoretical Division, CH 1211 Geneva, Switzerland}
\end{center}
\vfill
\begin{center} {\bf Abstract}
\end{center}
{\small
We generalize central--charge relations and differential identities of
$N=2$ Special Geometry to $N$ extended supergravity in any dimension $4 \leq D <10$,
and $p$--extended objects.

We study the extremization of the ADM mass per unit of $p$--volume of BPS
extended objects.
Runaway solutions for a ``dilaton'' degree of freedom leading to a vanishing
result are interpreted as BPS extremal states having vanishing Bekenstein--Hawking Entropy.
}
\vspace{2mm} \vfill \hrule width 3.cm
{\footnotesize
\noindent
$^*$ Supported in part by DOE grant
DE-FGO3-91ER40662, Task C.
and by EEC Science Program SC1*CT92-0789.}
\end{titlepage}
\section{introduction}
In recent time attempts to study non perturbative properties of gauge
theories \cite{sw} and string theories \cite{ht}, \cite{wit} have made an essential use of low energy
effective lagrangians incorporating the global and local symmetries
of the fundamental theories.
\par
In this analysis BPS states play an important role \cite{sen, ss}, especially in
connection with enhancement of gauge symmetries \cite{cdfv, ht2, w2} and more generally
for phase transitions which may be signaled by some BPS state
becoming massless at some point of the underlying moduli space.

Many of these phenomena can be studied, to some extent, by properties
of the effective supergravity theories and their central extensions
\cite{df}--\cite{ach},
which are the analogue of the non linear chiral lagrangians for QCD.
\par
The BPS states often appear as solitonic solutions of the
supergravity field equations in backgrounds preserving some of the
supersymmetries depending on the degree of extremality of the
solitonic state (see for instance \cite{gh}--\cite{dfkr}).
\par
Recently a lot of information on black holes and black $p$-branes in
diverse dimensions have been obtained using these methods
\cite{dlp}--\cite{dlr}.

For $N=2$,  $N=4$ and $N=8$ black holes in $D=4,5$ an almost complete analysis
of their solitonic configurations has been given
\cite{dr}--\cite{ct} and partial results
for $D > 4$ theories, both for extremal and non extremal situations,
 are available \cite{dkl}--\cite{cy}, \cite{kt1}--\cite{dr}.
The underlying geometry of the moduli space plays a fundamental role
in finding these solutions since the ADM mass or, more
generally, the mass per
unit of $p$--volume for $p$--extended objects depends on the
asymptotic value of the moduli and some other physical quantities,
such as the classical determination of the Bekenstein - Hawking
entropy formula, are also related to properties of the moduli space
\cite{lw, sv}.
For instance, extremal black holes preserving one
 supersymmetry in $D=4$ and $5$ dimensions  have an entropy formula
obtained in a rather moduli independent way
by minimizing the ADM mass in the moduli space \cite{fk1}.
\par
These results heavily   rely on properties which connect space-time
supergravity  with the underlying moduli space of the theory.
For instance,
properties of $N=2$ extremal black holes at $D=4$ and $5$   depend merely on
 the underlying geometry of the moduli space.
\par
These considerations lead to a simple understanding of no hair theorems
and to the possibility of describing the physics of the black hole horizon in
terms of
an effective potential encoding the
thermodynamical properties of the system \cite{gkk}.\\
In view of several non perturbative dualities between different
kinds of theories, a
given theory is truly specified by the dimension of space time
in which it lives,
the number of unbroken supersymmetries and the massless matter content.

The aim of the present work is to
 further extend these results  generalizing some of these considerations to
higher $N$ supergravities in diverse dimensions. We  study differential identities
 between different kinds of charges and  establish  sum rules which generalize the ones
previously obtained for $N=2$ theories. We also  study the extremization
of the ADM mass
of several $p$--extended BPS states and draw some conclusions about the
Bekenstein--Hawking entropy formula.

An expanded version of the present paper, focussed on the relation between central and matter 
charges in extended supergravities in any dimensions, will appear in a forthcoming publication.
In particular it will be discussed the relation between $N=2$ Special Geometry 
and the existence of a flat symplectic connection in all higher $N$ theories at $D=4$ \cite{adf}.

\section{Differential identities and sum rules for central and matter charges}

With the exception of $D=4$, $N=1,2$ and $D=5$, $ N=2$  all supergravity
theories contain scalar fields whose kinetic Lagrangian is described
by $\sigma$--models of the form $G/H$.\\
Here $G$ is a non compact group acting as an isometry group on the
scalar manifold while $H$, the isotropy subgroup, is of the form:
\begin{equation}
H=H_{Aut} \otimes H_{matter}
\end{equation}
 $H_{Aut}$ being the automorphism group of the supersymmetry algebra
 while $H_{matter}$ is related to the matter multiplets.
 (Of course $H_{matter}=\bfone$ in all cases where supersymmetric matter
 doesn't exist, namely $N>4$ in $D=4,5$ and in general in all
 maximally extended supergravities)

The coset manifolds $G/H$ and the automorphism groups for various
 supergravity theories for any $D$ and $N$  can be found in the literature
 (see for instance the reference book
\cite{sase}).
 As it is well known, the group G acts linearly on the $n=p+2$--forms
 field strengths $H^\Lambda _{a_1\cdots a_n}$ corresponding to the
 various $p+1$--forms appearing in the gravitational and matter
 multiplets. Here and in the following the index $\Lambda$ runs over
 the dimensions of some representation of the duality group $G$.

The true duality symmetry, acting on integral  quantized electric
and magnetics charges,
 is  the restriction of  the continuous group $G$ to the integers
 \cite{ht}.
 \par
 All the properties of the given supergravity theories for fixed $D$
 and $N$ are completely fixed in terms of the geometry of $G/H$
 namely in terms of the coset representatives $L$ satisfying the
 relation
 \begin{equation}
g L(\phi) = L(\phi ^\prime) h^{-1} (g,\phi)
\end{equation}
where $g\in G$, $h\in H$  and $\phi ^\prime =   \phi ^\prime
 (\phi)$,
 $\phi$ being the coordinates of $G/H$.
 In particular, as explained in the following, the kinetic metric for
 the $p+2$ forms $H^\Lambda$ is fixed in terms of $L$ and the
 physical field strengths of the interacting theories are "dressed"
 with scalar fields in terms of the coset representatives.

 This allows us to write down the central charges associated to the
 $p+1$-- forms in the gravitational multiplet in a neat way in terms
 of the geometrical structures of the moduli space.

In an analogous way also the matter $p+1$--forms of the matter
multiplets give rise to charges which, as we will see, are closely
related to the central charges. Note that when $p>1$ these central
charges do not appear in the usual supersymmetry algebra, but in the
extended version of it containing central generators $Z_{a_1 \cdots
a_p}$ associated to $p$--dimensional extended objects ($a_1 \cdots
a_p$ are a set of space--time antisymmetric
Lorentz indices) \cite{df, vpvh, tow,
ach}
\par
Our main goal is to write down the explicit form of the dressed
charges and to find relations among them analogous to those worked
out in $D=4$, $N=2$ case  by means of the Special Geometry relations \cite{cdf}\cite{cdfv}.
\par
To any $p+2$--form $H^\Lambda$ we may associate a magnetic charge ($D-p-4$--brane)
and  an
electric ($p$--brane) charge given respectively by:
\begin{equation}
g^\Lambda = \int _{S^{p+2}} H^\Lambda
\qquad \qquad
e_\Lambda = \int _ {S^{D-p-2}}  \cG _\Lambda
\end{equation}
where $\cG_{\Lambda}= {\partial \cL \over \partial H^\Lambda}$.
\par
These charges however are not the physical charges of the interacting
theory; the latter ones can be computed by looking at the
transformation laws of the fermion fields, where the physical
field--strengths appear dressed with the scalar fields.

Let us first introduce the central charges:
they are associated to the dressed $p+2$--forms $H^\Lambda$ appearing
in the supersymmetry transformation law of the gravitino 1-form.
Quite generally we have, for any $D$ and $N$:
\begin{equation}
\delta \psi_A = D\epsilon_A + \sum_{i} c_i L_{\Lambda_i AB} (\phi)
H^{\Lambda_i} _ {a_1\cdots a_{n_i}}\Delta^{a a_1\cdots a_n}
\epsilon^B V_a+ \cdots
\label{tragra}
\end{equation}
where:
 \begin{equation}
\Delta_{a a_1\cdots a_n}=\left( \Gamma _{a a_1 \cdots
a_{n_i}} - {n \over n-1} (D-n-1)\delta^a_{[a_1} \Gamma_{a_2\cdots
a_{n_i}]} \right).
\end{equation}
 Here $c_i$ are coefficients fixed by supersymmetry, $V^a$ is the
 space--time vielbein, $A=1,\cdots,N$ is the index acted on by the
 automorphism group, $\Gamma_{a_1\cdots a_n}$ are $\gamma$--matrices
 in the appropriate dimensions, and the sum runs over all the $p+2$--forms
 appearing in the gravitational multiplet. Here and in the following
 the dots denote trilinear fermion terms. $L_{\Lambda AB}$ is given
 in terms of the coset representative matrix of $G$.
 Actually it coincides with a subset of the columns of this matrix
 except in $D=4$ ($N>1$) and the for maximally extended
 $D=6,8$ supergravities since in those cases we have the slight
 complication  that the action of $G$ on the $p+2 =D/2$--forms is
 realized through the embedding of $G$ in $Sp(2n,\IR)$ or $O(n,n)$
 groups.\\
 Excluding for the moment these latter cases,
$L_{\Lambda AB}$  is actually a set of columnes of the (inverse) coset
representative  $L$ of $G$.
Indeed, let us decompose the representative of $G/H$ as follows:
\begin{equation}
L=(L^\Lambda_{\ AB}, L^\Lambda_{\ I}) \qquad\quad L^{-1} =(L^{AB}_{\
\
\Lambda}, L^I_{\ \Lambda})
\label{defl}
\end{equation}
where the couple of indices $AB$ transform as a symmetric tensor
under $H_{Aut}$
and $I$ is an index in the fundamental
representation of $H_{matter}$ which in general is an orthogonal group
(in absence of matter multiplets
$L\equiv ( L^\Lambda_{\ AB})$).
Quite generally we have:
\begin{equation}
 L_{AB\Lambda}L^{AB}_{\ \ \Sigma} - L_{I\Lambda}L^{I}_{\ \Sigma}=
\cN_{\Lambda\Sigma}\label{ndil}
\end{equation}
where
  $\cN$ defines the kinetic matrix of the $(p+2)$--forms $H^\Lambda$
 and the indices of $H_{Aut}$ and $H_{matter}$ (generally given by a  pseudoorthogonal group) are raised
 and lowered with the appropriate
metric in the given representation.
For maximally extended supergravities
 $\, \cN_{\Lambda\Sigma} = L_{AB\Lambda}L^{AB}_{\ \ \Sigma}$.

When $G$ contains an orthogonal factor $O(m,n)$, what happens
 for matter coupled supergravities in $D=5,7,8,9$, where $G=O(10-D,n)  \times O(1,1)$
 and in all the matter coupled $D=6$ theories,
 the coset
representatives of the orthogonal group satisfy: 
\begin{eqnarray}
L^t\eta L = \eta &\to &  L_{A\Lambda}L^{A}_{\ \ \Sigma} +  L_{I\Lambda}L^{I}_{\ \Sigma}=
\eta_{\Lambda\Sigma} \label{etadil}\\
L^t L = \cN &\to & L_{A\Lambda}L^{A}_{\ \ \Sigma} -  L_{I\Lambda}L^{I}_{\ \Sigma}=
\cN_{\Lambda\Sigma} \label{ndil1}
\end{eqnarray}
 where $\eta=\pmatrix{\bfone_{m\times m} & 0 \cr 0 & -\bfone_{n\times n}\cr}$
 is the $O(m,n)$ invariant metric  and $A=1,\cdots,m$; $I = 1,\cdots,n$.
(In particular,  setting the matter to zero, we have in these cases
$\cN_{\Lambda\Sigma}= \eta_{\Lambda\Sigma}$).\\

Note that both for matter coupled and maximally extended
supergravities we have:
\begin{equation}
L_{\Lambda AB} = \cN_{\Lambda\Sigma}L^{\Sigma}_{\ AB}
\end{equation}
From equation (\ref{tragra}) we see that the dressed
graviphoton $n_i$--forms field strengths are:
\begin{equation}
 T^{(i)}_{AB} = L_{\Lambda_i AB} (\phi) H^ {\Lambda_i}
\end{equation}
The magnetic central charges for BPS saturated $D-p-4$--branes
can be now defined
(modulo numerical factors to be fixed in each theory) by integration
of the dressed
field strengths as follows:
\begin{equation}
Z^{(i)}_{(m) AB} = \int _{S^{p+2}}L_{\Lambda_i AB}(\phi) H^{\Lambda_i}=
 L_{\Lambda_i AB}(\phi_0) g^{\Lambda_i}
 \label{carma}
\end{equation}
where $\phi_0$ denote the $v.e.v.$ of the scalar fields, namely
$\phi_0 = \phi(\infty)$ in a given background.

The corresponding electric central charges are:
 \begin{equation}
Z^{(i)}_{(e) AB} = \int _{S^{D-p-2}}L_{\Lambda_i AB}(\phi)
^{\ \star}H^{\Lambda_i}= \int _{S^{D-p-2}} \cN _{\Lambda_i\Sigma_i}
L^{\Lambda_i}_{\  AB}(\phi)
^{\ \star}H^{\Sigma_i}=
 L^{\Lambda_i}_{\ AB}(\phi_0) e_{\Lambda_i}
\end{equation}
 These formulae make it explicit that $L^\Lambda_{\ AB}$ and
 $L_{\Lambda AB}$ are related by electric--magnetic duality via the
 kinetic matrix.
 \par
 Note that the same field strengths (the graviphotons) which appear in the gravitino
 transformation laws are also present in the dilatino transformation laws
  in the following way:
  \begin{equation}
\delta \chi_{ABC} = \cdots  +
\sum_{i} b_i L_{\Lambda_i AB} (\phi)
H^{\Lambda_i} _ {a_1\cdots a_{n_i}}\Gamma^{ a_1\cdots a_{n_i}} \epsilon_C + \cdots
\label{tradil}
\end{equation}

 In an analogous way, when vector multiplets are present,
 the matter vector field
 strengths are dressed with the columns $L_{\Lambda I}$
 of the coset element (\ref{defl})
 and they
 appear in the transformation laws of the gaugino fields:
  \begin{equation}
\delta \lambda^I_{A} = \Gamma^a P^I_{AB , i}
\partial_a \phi^i \epsilon^B +
 c L_{\Lambda}^{\ I} (\phi)
F^{\Lambda} _ {ab}\Gamma^{ ab} \epsilon_A  + \cdots
\label{tragau}
\end{equation}
  where  $ P^I_{AB , i}$ is the vielbein of the coset manifold
  spanned by the scalar fields of the vector multiplets and $c$ is a
  constant fixed by supersymmetry (in $D=6$, $N=(2,0)$ and $N=(4,0)$ the
  2--form $F^{\Lambda}_{ab} \Gamma^{ab}$ is replaced by the 3--form
  $F^{\Lambda}_{abc} \Gamma^{abc}$).

In the same way as for central charges, one finds the
magnetic matter charges:
\begin{equation}
  Z_{(m) A} ^{\ I} = \int _{S^{p+2}} L_\Lambda^{\ I} F^\Lambda
   =   L_\Lambda^{\ I} (\phi_0) g^\Lambda
\end{equation}
while the electric matter charges are:
\begin{equation}
Z_{(e) I} = \int _{S^{D-p-2}}L_{\Lambda I}(\phi)
^{\ \star}F^{\Lambda}= \int _{S^{D-p-2}} \cN _{\Lambda\Sigma}
L^{\Lambda}_{\  I}(\phi)
^{\ \star}F^\Sigma =
 L^{\Lambda}_{\ I}(\phi_0) e_{\Lambda}
\end{equation}
 \par
 The important fact to note is that the central charges and matter
 charges satisfy relations and sum rules analogous to those derived
 in $D=4$, $N=2$ using Special Geometry techniques \cite{cdf}.
 In the general case these sum rules are inherited from the
 properties of the coset manifolds $G/H$, namely from the
 differential and algebraic properties satisfied by the coset
 representatives $L^\Lambda_{\ \Sigma}$.

Indeed, for a general coset manifold we may introduce the
left--invariant 1--form $\Omega=L^{-1} d L$ satisfying the
relation (see for instance \cite{cadafr}):
\begin{equation}
d \Omega + \Omega \wedge \Omega =0
\label{mc}
\end{equation}
 where \begin{equation}
\Omega=\omega^i T_i + P^\alpha T_\alpha
\label{defomega}
\end{equation}
  $T_i, T_\alpha$ being the generators of $G$ belonging respectively to the Lie
  subalgebra $\IH$ and to the coset space algebra $\IK$ in the
  decomposition
  \begin{equation}
\IG = \IH + \IK
\label{hk}
\end{equation}
 $\IG$ being the Lie algebra of $G$. Here $\omega^i$ is the $\IH$
 connection and $P^\alpha$  is the vielbein of $G/H$.

Since in all the cases we will consider $G/H$ is a symmetric space
($[\IK , \IK ] \subset \IH$), $\omega^i C_i^{\ \alpha\beta}$
($C_i^{\ \alpha\beta}$ being the structure constants of $G$)
can be identified
with the Riemannian spin connection of $G/H$.
\par
Suppose now we have a matter coupled theory. Then, using
the decomposition (\ref{hk}), from
(\ref{mc}) and (\ref{defomega}) we get:
\begin{equation}
dL^\Lambda_{\ AB} = L^\Lambda_{\ CD} \omega^{CD}_{\ \ AB} +
L^\Lambda_{\ I} P^I_{AB}
\label{cosetmc}
\end{equation}
where $P^I_{AB}$ is the vielbein on $G/H$ and
 $\omega^{CD}_{\ \ AB}$ is the $\IH$--connection in the given
 representation.
It follows:
\begin{equation}
D^{(H)} L^\Lambda_{\ AB}= L^\Lambda_{\ I} P^I_{AB}
\label{dllp}
\end{equation}
where the derivative is covariant with  respect to the
$\IH$--connection  $\omega^{CD}_{\ \ AB}$.
Using the definition of the magnetic dressed charges given in
(\ref{carma}) we obtain:
\begin{equation}
D^{(H)} Z_{AB}= Z_{I} P^I_{AB}
\label{dz}
\end{equation}
 This is a prototype of the formulae one can derive in the various
 cases for matter coupled supergravities.
 To illustrate one possible application of this kind of formulae let
 us suppose that in a given background preserving some number of
 supersymmetries $Z_I=0$ as a consequence of $\delta\lambda^I_A=0$.
 Then we find:
 \begin{equation}
D^{(H)} Z_{AB}=0 \to d(Z_{AB} Z^{AB} )=0
\end{equation}
 that is the square of the
 central charge reaches an extremum with respect to the
 $v.e.v.$ of the moduli fields.

 For the maximally extended supergravities there are no matter
 field--strengths and the previous differential relations become
differential relations between central charges only.
Indeed in this case the Maurer--Cartan equations become:
\begin{equation}
dL^{\Lambda}_{\ AB} = L^{\Lambda}_{\ CD} \Omega^{CD}_{\ \ AB}+
L^{\Lambda}_{\ CD}  P^{CD}_{\ \ AB}
\end{equation}
 where now $AB$ runs over the same set of values as $\Lambda$.
 Therefore we get:
  \begin{equation}
D^{(H)} L^\Lambda_{\ AB}= L^\Lambda_{\ CD} P^{CD}_{\ \ AB}
\end{equation}
   that is:
  \begin{equation}
D^{(H)} Z_{AB}= Z^{CD} P_{CD AB}
\end{equation}
 This relation implies that the vanishing of a subset of central
 charges forces the vanishing of the covariant derivatives of some
 other subset.
 Typically, this happens in some
 supersymmetry preserving backgrounds where
 the requirement $\delta\chi_{ABC}=0$ corresponds to the vanishing of
 just a subset of central charges.

 Finally, from the coset representatives relations
 (\ref{ndil}) (\ref{etadil}) it
 is immediate to obtain sum rules for the central and matter charges
 which are
the counterpart of those found in $N=2$, $D=4$ case using Special
Geometry.
 Indeed, let us suppose e.g. that the group $G$ is
 $G=O(10-D,n)\times O(1,1)$, as it
 happens in general for all the minimally extended supergravities in
 $7 \leq D \leq 9$,  $D=6$ type $IIA$ and
 $D=5$, $N=2$.
  The coset representative is now a tensor product $L \to e^\sigma L$, where
  $e^\sigma$ parametrizes the $O(1,1)$ factor.\\
  We have, from   (\ref{etadil})
 \begin{equation}
L^t \eta L =\eta
\end{equation}
 where $\eta$ is the invariant metric of $O(10-D,n)$ and  from  (\ref{ndil})
 \begin{equation}
e^{-2\sigma}(L^{t} L)_{\Lambda\Sigma} =\cN_{\Lambda\Sigma}.
\end{equation}
 Using the decomposition (\ref{hk}) one finds:
 \begin{equation}
Z_{AB} Z^{AB} + Z_I Z^I = g^\Lambda \eta _{\Lambda\Sigma} g^\Sigma
e^{-2\sigma}
\end{equation}
   \begin{equation}
Z_{AB} Z^{AB} - Z_I Z^I = g^\Lambda \cN _{\Lambda\Sigma} g^\Sigma
\end{equation}
 In more general cases analogous relations of the same kind can be
 derived.
 \vskip5mm
 \par
Let us now see how these considerations modify in  the case of extended objects which
can be dyonic, i.e. for $p=(D-4)/2$.
Following Gaillard and Zumino \cite{gz}, for $p$ even ($D$ multiple of 4)
 the duality group
$G$ must have a symplectic embedding in $Sp(2n,\IR)$;
for $p$ odd ($D$ odd multiple of 2),
the duality group is always $O(n,m)$ where n,m are respectively the number of
self--dual and anti self--dual $p+2$--forms.
 \\
 In $D=4$, $N>2$ we may decompose the vector field--strengths in self--dual and
 anti self--dual parts:
 \begin{equation}
\cF^{\mp} = {1\over 2}(\cF\mp {\rm i} ^{\ \star} \cF)
\end{equation}
  According to the Gaillard--Zumino construction, $G$ acts on the
  vector $(\cF^{- \Lambda},\cG^{-}_\Lambda)$
  (or its complex conjugate) as a subgroup of
  $Sp(2 n_v,\IR)$ ($n_v$ is the number of vector fields)
with duality transformations interchanging electric and magnetic
 field--strengths:
 \begin{equation}
{\cal S}
\left(\matrix {\cF^{-\Lambda} \cr
\cG^-_\Lambda\cr}\right)=
\left(\matrix {\cF^{-\Lambda} \cr
\cG^-_\Lambda\cr}\right)^\prime
\end{equation}
 where:
 \begin{eqnarray}
\cG^-_\Lambda&=&\bar \cN_{\Lambda\Sigma}\cF^{-\Sigma}\\
 \cG^+_\Lambda&=& \cN_{\Lambda\Sigma}\cF^{+\Sigma}
\end{eqnarray}
\begin{equation}
 {\cal S}=\left( \matrix{A& B\cr C & D \cr}\right)\in G \subset Sp(2 n_v,\IR)
\end{equation}
  and $\cN_{\Lambda\Sigma}$, is the matrix appearing in the kinetic
  part of the vector Lagrangian:
  \begin{equation}
  \cL_{kin}= {\rm i}\bar \cN_{\Lambda\Sigma}\cF^{-\Lambda} \cF^{-\Sigma} + h.
  c.
  \end{equation}

Using a complex basis in the vector space of $Sp(2 n_v)$, we may
rewrite the
symplectic matrix in the following way:
\begin{equation}
{\cal S}
\to U=\left(\matrix{\phi_0 & \bar \phi_1 \cr
\phi_1 & \bar \phi_0 \cr}\right)
\end{equation}
where:
\begin{eqnarray}
 \phi_0 &=& {1 \over 2} (A-{\rm i} B ) + {{\rm i} \over 2} (C - {\rm i} D)
 \\
 \phi_1 &=& {1 \over 2} (A-{\rm i} B ) - {{\rm i} \over 2} (C - {\rm i} D)
\end{eqnarray}
Defining:
   \begin{eqnarray}
  f^\Lambda&=& -{1 \over
  \sqrt{2}}(\phi_0 + \phi_1)\\
  h_\Lambda &=& -{1 \over
  \sqrt{2}}(\phi_0 - \phi_1).
\end{eqnarray}
 $f$ and $h$  are coset representatives of $G$ embedded in
 $Sp(2 n_v, \IR)$ and can be constructed in terms of the $L$'s.
The kinetic matrix $\cN$ turns out to be:
\begin{equation}
\cN= hf^{-1}
\label{nfh-1}
\end{equation}
 and transforms projectively under duality rotations:
 \begin{equation}
\cN^\prime = (C+ D \cN) (A+B\cN)^{-1}
\end{equation}

The requirement $ {\cal S} \in Sp(2 n_v, \IR)$ implies:
 \begin{equation}
\left\lbrace\matrix{{\rm i}(f^\dagger h - h^\dagger f) &=& \bfone \cr
(f^\dagger \bar h - h^\dagger \bar f) &=& 0\cr} \right.
\label{specdef}
\end{equation}
  By using (\ref{nfh-1}) we find that
   \begin{equation}
   (f^t)^{-1} = {\rm i} (\cN - \bar \cN)\bar f
\end{equation}
 As a consequence, in the transformation law of gravitino (\ref{tragra})
 and gaugino (\ref{tragau})
 we have to substitute
 \begin{equation}
(L_{\Lambda AB}, L_{\Lambda I}) \to  (f_{\Lambda AB}, f_{\Lambda I})
\end{equation}
In particular, the dressed graviphotons and matter vectors take the
symplectic invariant form:
\begin{eqnarray}
T^-_{AB}&=& f^\Lambda_{AB}(\cN - \bar \cN)_{\Lambda\Sigma}\cF^{-\Sigma}=
f^\Lambda_{AB} \cG^-_\Lambda - h_{\Lambda AB} \cF^{-\Lambda}  \\
  T^-_{I}&=& f^\Lambda_{I}(\cN - \bar \cN)_{\Lambda\Sigma}\cF^{-\Sigma}=
f^\Lambda_{I} \cG^-_\Lambda - h_{\Lambda I} \cF^{-\Lambda}
\end{eqnarray}
The corresponding central and matter charges become:
\begin{eqnarray}
Z_{AB}&=&
f^\Lambda_{AB} e_\Lambda - h_{\Lambda AB} g^{\Lambda}
\label{central}\\
  Z_{I}&=&
f^\Lambda_{I} e_\Lambda - h_{\Lambda I} g^{\Lambda}
\end{eqnarray}
 We see that the presence of dyons in $D=4$ is related to the
 symplectic embedding.
Also in this case one can obtain differential relations
and a sum rule among the charges.
The sum rule has the following form:
\begin{equation}
Z_{AB} \bar Z_{AB} + Z_I \bar Z_I = -{1\over 2} P^t \cM (\cN) P
\end{equation}
where $\cM(\cN)$ and $P$ are:
\begin{equation}
\cM = \left( \matrix{ \bfone & 0 \cr - Re \cN &\bfone\cr}\right)
\left( \matrix{ Im \cN & 0 \cr 0 &Im \cN^{-1}\cr}\right)
\left( \matrix{ \bfone & - Re \cN \cr 0 & \bfone \cr}\right)
\label{m+}
\end{equation}
\begin{equation}
P=\left(\matrix{g^\Lambda \cr e_ \Lambda \cr} \right)
\label{eg}
\end{equation}
\par
Furthermore the Maurer--Cartan equations (\ref{dllp}) for the coset
representatives of $G/H$
imply analogous Maurer--Cartan equations for the embedding coset
representatives $(f,h)$:
\begin{equation}
\nabla^{(\IH)} (f,h) = (\bar f, \bar h)P^{(\IK)}
\end{equation}
The differential relations among central and matter charges and
 their sum rules can then be found in a way analogous to that shown
 before for the odd dimensional cases.
 \par
  For $D=8$, $N=2$ the situation is exactly similar to the $D=4$ case,
  where the 2--forms field--strengths are to be understood as  4--forms.
 In the case at hand the 4--form in the gravitational multiplet
and its dual are a doublet under the duality group $Sl(2,\IR)$.
 \par
Finally in $D=6$
the  3--form field strengths $H^\Lambda$ which appear in the
gravitational and/or tensor multiplet have a definite self--duality
 \begin{equation}
H^{\pm \Lambda} = {1\over 2} (H^\Lambda \pm ^\star H^\Lambda)
\end{equation}
  In this case the duality group is of the form $G=O(m,n)$.
  Except for the left--right symmetric cases $N=(4,4)$ and $N=(2,2)$, the number of
  self--dual tensors $H^{+\Lambda_1}$ in the gravitational multiplet,
  $\Lambda_1 =1,\cdots m$
 and antiself--dual tensors $H^{-\Lambda_2} $ in the matter
 multiplet, $\Lambda_2 =1,\cdots n$
 are different in general and $G$
  acts in its fundamental representation on
  $(H^{+\Lambda_1}, H^{-\Lambda_2})$ so that no embedding is required.

The procedure to find the charges and their relations is thus
completely analogous to the odd dimensional case, that is
\begin{equation}
Z_{AB} = L_{\Lambda AB} g^\Lambda
\end{equation}
 However, due to the
relation:
  \begin{equation}
\cN_{\Lambda\Sigma} ^{\ \ \star}H^\Lambda = \eta_{\Lambda\Sigma} H^\Sigma,
\label{elmagn}
\end{equation}
where $\eta$ and $\cN$ are defined in terms of the coset representatives of ${O(n,m)\over O(m)\times O(n)}$
as in
(\ref{etadil}), (\ref{ndil}),
we have no distinction among electric and magnetic charges. Indeed
\begin{equation}
e_\Lambda= \int \cN_{\Lambda\Sigma} ^{\ \ \star}H^\Lambda =
\int  \eta_{\Lambda\Sigma} H^\Sigma =\eta_{\Lambda\Sigma} g^\Sigma
\end{equation}
 For the maximally extended case, we have an equal number (5)
 of self--dual and anti self--dual field strengths
 and therefore a Lagrangian
exists.
 The group
 $G=O(5,5)$ rotates among themselves $H^+$ and $H^-$ in the
 representation $\underline {10}$.
 The analogous of the Gaillard--Zumino construction in this case would
 define an $O(5,5)$ embedding of $O(5)$ rotating among themselves
 $H^+ ,\cG^+$ or $H^-, \cG^-$ where
 \begin{equation}
\cG^\pm=\cN_\pm H^\pm
\end{equation}
  where $\cN_-=-(\cN_+)^t$ is the kinetic metric of the tensors in the
  Lagrangian.
In this case we obtain a formula analogous to (\ref{central})
 which is however invariant under $O(5,5)$ instead than $Usp(n,n)$:
\begin{equation}
Z_{\pm AB} = f^\Lambda_\pm e_\Lambda + h_{\Lambda \pm}g^\Lambda
\label{z+_}
\end{equation}
\section{Considerations on maximally and non maximally extended supergravities}
We now consider some properties and differences of
 maximally extended theories versus non maximal ones.
 $D=4,5$ maximally extended theories ($N=8$) with solutions preserving one supersymmetry
have been studied in
 ref.
 \cite{kako, fk1} and will not be discussed further.
The Bekenstein--Hawking entropy is expressed in terms of the quartic
and cubic invariant \cite{cj}
of $E_{7(7)}$ \cite{kako} and $ E_{6(6)}$ \cite{fk1} respectively.
We give here a simple proof of why this is the case.
The matrix $L$ defining the coset representative transforms under
$G_L\times G_R$ as $L \to g_L L g_R^{-1}$,
The central charge matrix ( with its complex conjugate) is
a vector $Z=LP$  under $G_R$ where $P=(e,g)$ is a vector under $G_L$.
therefore any $G_R$--invariant $I$  constructed out of $Z$ is
independent of $L$:
  $I(Z)=I(LP)= I(P)$.

We now consider maximal supergravities for $D>5$;
for $5\leq D \leq 7$ we have only $p$=0,1--branes (with their duals
$p^\prime = 0,1,2,3$) while for $D=8,9$
also $p=2$--branes occur (together with their duals $p^\prime =2,3,4,5$)
Here we report as an example how the coset representatives
spell out in $D=9$. The duality group is, in this case,
$Sl(2,\IR)\times O(1,1)$ while the coset manifold is \cite{sase}:
 \begin{equation}
G/H = {  Sl(2, \IR) \over O(2)} \times O(1,1)
\end{equation}
and the field content and  group assignements are given in Table 1,
where $A,B,C$ are $O(2)$ vector indices, $L^\Lambda_{\ AB}$ is the
coset representative of ${Sl(2, \IR) \over O(2)}$ symmetric and
traceless in $A,B$, $e^\sigma$ parametrizes $O(1,1)$, $\Lambda =1,2$
are indices of $Sl(2, \IR)$ in the defining representation, $\chi
_{ABC}$ is completely symmetric and can be decomposed as
\begin{equation}
\chi_{ABC} = \buildrel \circ \over \chi _{ABC} + \delta_{(AB} \chi _{C)}
\end{equation}
From the analysis of the fermions transformation laws we get
the following  magnetic central charges:
\begin{eqnarray}
 Z^{(4)} &=& e^{-\sigma} g   \qquad g= \int H^{(4)} \\
 Z^{(3)}_{AB} &=& e^{-\sigma}L_{\Lambda AB}
 g^{\Lambda}  \qquad
 g^\Lambda= \int H^{(3)\Lambda} \\
Z^{(2)}_{(AB)} &=&  L_{\Lambda AB}
m^{\Lambda}     \qquad
 m^\Lambda= \int F^{(2)\Lambda}\\
Z^{(2)} &=& e^{-\sigma} m   \qquad
 m= \int F^{(2)}
\end{eqnarray}
where the superscript in the field--strengths denotes their order as forms.

Using now the Maurer--Cartan equations for the coset representative
$e^{-\sigma} L_{\Lambda AB}$ we find:
\begin{equation}
\nabla^{O(2)} (e^{-\sigma}L_{\Lambda AB}) = e^{-\sigma}
( L_{\Lambda AC} P_{CB} - d\sigma L_{\Lambda AB} )
\end{equation}
   Therefore:
   \begin{eqnarray}
\partial _\sigma \left( \matrix{ Z^{(4)} \cr Z^{(3)}_{AB} \cr Z^{(2)} }
\right) &=& - \left( \matrix{ Z^{(4)} \cr Z^{(3)}_{AB} \cr Z^{(2)} }
\right) \label{dsigmaz}\\ \nonumber\\
\nabla_i \left( \matrix{ Z^{(3)}_{AB} \cr Z^{(2)}_{AB} }
\right) &=&  \left( \matrix{ Z^{(3)}_{AC} \cr Z^{(2)}_{AC} }
\right)P_{CB,i}\label{diz}
\end{eqnarray}
From eqs. (\ref{dsigmaz}), (\ref{diz}) we see that the extremization
of the central charges of the singlet 0--brane, the doublet of 1--branes,
 and the singlet  2--brane occur at zero value of the central charge.
 This corresponds to a minimum with runaway
behaviour of the $O(1,1)$ dilatonic field at $\sigma =\infty$.
We can conclude in this case that for this $p$--extended object
there is necessarily
a zero of the area--entropy formula.
Similar conclusions do not follow immediately for maximally
extended theories at $D<9$ where the duality group
does not contain such a  $O(1,1)$ factor.

If we  consider instead all non maximal theories with
16 supersymmetries (which reduce to $N=4$ in 4 and 5 dimensions)
they all have a $O(1,1)$ factor in the duality group
(with the exception of chiral $(4,0)$ theory  in $D=6$,
which does not have vectors altogether).
The vectors are in the fundamental of the $O(10-D, n)$
T--duality group and are also charged
  under the S--duality group $O(1,1)$ which reduces to $\ZZ_2$ when
  it is
  restricted to the integers.
Therefore, by the same argument as before, one can prove
that the extremal 0--branes
have only the runaway solution $\sigma= \infty$ as an extremum.
As a consequence  all 0--branes in this theories
 have vanishing area--entropy formula at the extremum.

\begin{table}
\begin{center}
\caption{\sl transformation properties of fields in D=9, N=2}
\label{tab1}
\begin{tabular}[h]{|c||c|c|c|c|c|c|c|c|c|}
\hline
 D=9, N=2&$ V^a_\mu $& $C_{\mu\nu\rho}$ & $B^\Lambda_{\mu\nu}$
&$A^{\Lambda}_\mu$& $A_\mu $&
$L^\Lambda_{\ AB}$& $e^\sigma$& $ \psi^A_\mu $ & $ \chi^{ABC}$ \\
\hline
\hline
 $Sl(2, \IR)$ & 1 & 1 &2 & 2 & 1 & 2 & 1 & 1 & 1 \\
\hline
$O(2)$ & 1 & 1 & 1 & 1 & 1 & 2 & 1 & 2 & 2 + 2 \\
\hline
$O(1,1)$ & 0 & 1 & 1 & 0 & 1 & 0 & 1 & 0 & 0 \\
\hline
\end{tabular}
\end{center}
\end{table}

We incidentally note that the $D=5$ case is exceptional
in this respect due to the fact that in top of the
$10 -D+ n$ vectors having the same $O(1,1)$ charges there is an extra
 vector singlet (dual to $B_{\mu\nu}$) with different $O(1,1)$ charge,
 whose virtue is precisely to stabilize the $O(1,1)$
 mode and to give a finite non zero extremized ADM mass.

 We can rephrase the previous results in a more group--theoretical
 setting: since the extremized ADM mass is $G$ invariant, it may only
 depend on $G$ invariant quantities constructed with charged
 representations of $G$.
 As $G$ contains a $O(1,1)$ factor, an invariant is possible only
 if the charges carry different $O(1,1)$ quantum numbers. This
 happens only in $D=5$, $N=4$, but not for $D>5$. Indeed, in $D=5$, the
 vectors are in the reducible representation $(-2,1) \oplus (1,5+n)$
 of $O(1,1)\times O(5,n)$, where the $O(1,1)$ quantum number denotes
 the scaling properties of fields under $\sigma \to \sigma+c$.

This argument can actually be extended to more general cases, where
the $O(1,1)$ factor is not present, by using group--theoretical
arguments on the $G$--representatives of the charges.
In fact, if the $G$--representation of the charges does not admit an
invariant, either the extremum doesn't exist
or the extremized mass can not depend on the charges.
Inspection of the representation content shows that this is indeed
the case for any $p$--branes in $D>6$.

An exception is $D=6$, for which invariants exist for $p=1$ BPS
states (for $(2,0)$, $(4,0)$ chiral theories
and for $(4,4)$ maximal theory). However in the latter case the $p=0$
BPS states (black--holes) are in the spinor representation  of
$O(5,5)$ with no quadratic invariant. Hence there are not
black--holes with finite entropy at $D=6$, at least if we assume that
they can be obtained by decompactification of $D=5$ massless
black--holes corresponding to the chain $E_6 \to O(1,1) \times
O(5,5)$.

Note that in $D=4$ the invariant was possible because the $G$ group
is $Sl(2,\ZZ) \times O(6,n;\ZZ)$ with invariant $T_{\Lambda\Sigma}
T^{\Lambda\Sigma}$ where $T_{\Lambda\Sigma}$ is the $Sl(2,\IR)$
invariant skew tensor
\begin{equation}
T_{\Lambda\Sigma}=P_\Lambda Q_\Sigma - P_\Sigma Q_\Lambda
\end{equation}

\section{Conclusions}
In this note we have generalized previous results on central charges and their properties
to generic extended supergravities in $D$ dimensions and to generic BPS states describing extremal $p$--branes.

We have assumed  that the area per unit of $p$--brane volume is proportional to
    the BPS mass per unit of $p$--brane volume under the
 condition that a solution
  with one residual supersymmetry can be found \cite{bl,bbj,gkt,kklp}.
Under this condition we have been able to prove that in all theories with
 16 supersymmetries there are no
extremal 0--branes
with finite horizon area for $D > 5$.
Similar results are also obtained for the $D \geq 7$ maximally extended
theories
(with 32 supersymmetries).
If, on the other hand, solutions with finite entropy will be found,
this will imply that one of our hypotesis has been evaded.
BPS saturated 0--branes and 1--branes and their duals
are expected to have finite
area--entropy formula for $D=4,5,6$ respectively.

Finally it is worth noticing that our framework can also be applied to
the study of phase-transitions corresponding to a vanishing central
charge at some point of the moduli space.
For BPS saturated $p$--branes
this correspond to tensionless extended objects with infinitely many
point--particles becoming light \cite{wit2, sw2, dlp, kmv}.

 \section*{Acknowledgements}
 We would like to thank R. Kallosh and R. Khuri for helpful
discussions.

\end{document}